\documentstyle[epsfig, rotate, aps, pre, preprint]{revtex}
\begin{document}
\widetext

\title{A comparison between broad histogram and multicanonical methods}

\author{A. R. Lima\cite{arlima}, P. M. C. de Oliveira and T. J. P. Penna}
\address{Instituto de F\'{\i}sica, Universidade Federal Fluminense \\
  Av. Litor\^anea, s/n$^o$ - 24210-340 Niter\'oi, RJ, Brazil}

\date{\today}

\maketitle

\begin{abstract}
  We discuss the conceptual differences between the Broad Histogram
  (BHM) and reweighting methods in general, and particularly the
  so-called Multicanonical (MUCA) approaches. The main difference is
  that BHM is based on microcanonical, fixed-energy averages which
  depends only on the good statistics taken {\bf inside} each energy
  level. The detailed distribution of visits among different energy
  levels, determined by the particular dynamic rule one adopts, is
  irrelevant. Contrary to MUCA, where the results are extracted from
  the dynamic rule itself, within BHM any microcanonical dynamics could
  be adopted. As a numerical test, we have used both BHM and
  MUCA in order to obtain the spectral energy degeneracy of the
  Ising model in $4 \times 4 \times 4$ and $32 \times 32$ lattices,
  for which exact results are known. We discuss why BHM gives more
  accurate results than MUCA, even using {\bf the same} Markovian
  sequence of states. In addition, such advantage increases for larger
  systems.
  
\end{abstract}

\section{Introduction}
Development of tools for optimization of computer simulations is a
field of great interest and activity. Cluster updating algorithms
\cite{coniglio,swendsen92,wolff89}, probability reweighting procedures
\cite{salzburg59,dickman84,swendsen93} and, more recently, methods
\cite{berg91,lee93,hesselbo95} that obtain directly the spectral
degeneracy $g(E)$ are a few examples of very successful approaches
(for reviews of these methods see, for instance,
\cite{marinari96,defelicio96,newman99} and references therein). The
Multicanonical (MUCA) and Broad Histogram (BHM) methods belong to the
former category. The Entropic Sampling Method (ESM) \cite{lee93} was
proven to be an equivalent formulation of MUCA \cite{berg95}. From
the knowledge of $g(E)$, these methods allow us to obtain any
thermodynamical quantity of interest for the system under study, as
the canonical average
\begin{equation}
\label{canonical}
\langle Q\rangle_T = {\sum_E g(E) \langle Q(E)\rangle \exp(-E/T)
\over \sum_E g(E) \exp(-E/T)}
\end{equation}
of some macroscopic quantity $Q$ (magnetization, density,
correlations, etc). Both sums run over all allowed energies (for
continuous spectra, they must be replaced by integrals), $T$ is the
fixed temperature, and the Boltzmann constant was set to unity. The
degeneracy function $g(E)$ simply counts the number of states with
energy $E$ (which, must be interpreted as a density within a narrow
window ${\rm d}E$, for continuous spectra). Also,
\begin{equation}
\label{microcanonical}
\langle Q(E)\rangle = {\sum_{S[E]} Q_S \over g(E)}
\end{equation}
is the microcanonical, fixed-$E$ average of $Q$. The sum runs {\bf
uniformly} over all states $S$ with energy $E$ (or, again, within the
small window ${\rm d}E$, for continuous spectra). Note that neither
$g(E)$ nor $\langle Q(E)\rangle$ depend on the particular environment
the system is actually interacting with, for instance the canonical
heat bath represented by the exponential Boltzmann factors in eq.
(\ref{canonical}). Thus, these methods go far beyond the canonical
ensemble: once $g(E)$ and $\langle Q(E)\rangle$ were already
determined for a given system, one can study its behavior under
different environments, or different ensembles, using the same $g(E)$
and $\langle Q(E)\rangle$. Accordingly, as an additional advantage,
only one computer run is enough to evaluate the quantities of
interest in a large range of temperatures and other parameters.

MUCA was introduced in 1991 by Berg and Neuhaus\cite{berg91}. The
basic idea of the method is to sample microconfigurations performing
a biased random walk (RW) in the configuration space leading to
another unbiased random walk (i.e. uniform distribution) along the
energy axis. Thus, the visiting probability for each energy level $E$
is inversely proportional to $g(E)$. By tuning the acceptance
probability of movements in order to get a uniform distribution of
visits along the energy axis, one is able to get $g(E)$ at the end.
MUCA has been proven to be very useful and efficient to obtain
results in many different problems, such as first order phase
transitions, confinement/deconfinement phase transition in SU(3)
gauge theory, relaxation paths \cite{shteto97}, conformal studied of
peptides, helix-coil transition and protein folding, evolutionary
problems \cite{choi97} and to study phase equilibrium in binary lipid
bilayer \cite{besold99} (for reviews of the method see
\cite{mucareviews}).

BHM was introduced three years ago by de Oliveira {\em et al}
\cite{pmco96}. It is based on an exact relation between the spectral
degeneracy, $g(E)$, and the microcanonical averages of some special
macroscopic quantities. A remarkable feature of BHM is its
generality, since these macroscopic quantities can be averaged by
different procedures
\cite{pmco96,pmco98a,pmco98b,pmco98c,pmco98d,pmco99,wang98,munoz98,lima99,wang99,promb,j1j2}.
Because it is not restricted to a rule like a biased random walk,
more adequate dynamics can be adopted for each different application.
BHM has been applied to a variety of magnetic systems such as the 2D
and 3D Ising Models (also with external fields, next nearest neighbor
interactions), 2D and 3D XY and Heisenberg Models, Ising Spin Glass
\cite{pmco96,pmco98a,pmco98b,pmco98c,pmco98d,pmco99,wang98,munoz98,lima99,wang99,promb,j1j2}
with accurate results in a very efficient way. Theoretically, the
method can be applied to any statistical system \cite{pmco98c}.
Another distinguishing feature of BHM is that its numerical results
do not rely on the number of visits $H(E)$ to each energy level, a
quantity which is updated by one unit for each new visited state. On
the contrary, BHM is based on microcanonical averages of macroscopic
quantities. Each visited state contributes with a macroscopic upgrade
for the measured quantities. Thus the numerical accuracy is much
better than that obtained within all other methods, better yet for
larger and larger systems, considering the same computer effort.

Besides the practical points described above, the most important feature of
BHM is the following conceptual one. All reweighting methods
\cite{salzburg59,dickman84,swendsen93,berg91,lee93,hesselbo95} depend on the
final distribution of visits $H(E)$ along the energy axis.  Histogram methods
\cite{salzburg59,dickman84,swendsen93} adopt a canonical dynamics, getting
$H_{T_0}(E)$ for some fixed temperature $T_0$; a new distribution $H_T(E)$ is
then analytically inferred for another (not simulated) temperature $T$.
Following the same reasoning, one can also obtain $g(E)$ \cite{dickman84}.
Multicanonical approaches \cite{berg91,lee93,hesselbo95}, on the other hand,
tune appropriate dynamics in order to obtain a flat distribution $H(E)$.  In
both cases, the actually implemented transition probabilities from energy
level $E$ to another value $E'$ are crucial. In other words, in both cases the
results depend on the comparison of $H(E)$ with the neighboring $H(E')$. All
those reweighting methods are, thus, extremely sensitive to the particular
dynamic rule adopted during the computer run, i.e. to the prescribed
transition probabilities from $E$ to $E'$.

BHM is not a reweighting method. It does not perform any reweighting on the
distribution of visits $H(E)$. It needs only the knowledge of the
microcanonical, fixed-$E$ averages of some particular macroscopic quantities.
The possible transitions from the current energy $E$ to other values are {\bf
  exactly} taken into account within these quantities (see section
\ref{bh:section}), instead of performing a numerical measurement of the
corresponding probabilities during the computer run.  Thus, the only important
role played by the actually implemented dynamic rule is to provide a good
statistics {\bf within each energy level, separately}: the relative weight of
$H(E)$ as compared to $H(E')$, i.e.  the relative visitation frequency for
different energy levels, is completely irrelevant. One can even decide to
sample more states inside a particularly important region of the energy axis
(near the critical point, for instance) \cite{pmco99}, instead of a flat
distribution. In short, any dynamic rule can be adopted within BHM, the only
constraint is to sample with uniform probability the various states belonging
to the same energy level, not the relative probabilities concerning different
energies.

In this paper we have used the formulation of MUCA given by Lee \cite{lee93}
called Entropic Sampling, which, from now on, we call ESM.  We present a
comparison between ESM/MUCA and BHM, focusing on both accuracy and the use of
CPU time. We choose to start our study with the same example used in the
original ESM paper by Lee \cite{lee93}: the $4 \times 4 \times 4$ simple cubic
Ising model, for which the exact energy spectrum is known \cite{pearson82}.
Our results show that BHM gives more accurate results than ESM/MUCA with the
same number of Monte Carlo steps.  Despite the fact that one Monte Carlo step
in BHM takes more CPU time, in measuring further macroscopic averages, the
overall CPU time is smaller for the same accuracy, at least for this model.
Also, BHM can be applied to larger lattices without the problems faced by
ESM/MUCA, as we show in our simulations of the same model in a $32 \times 32$
lattice.

This paper is structured as follows: in section \ref{es:section} and
\ref{bh:section} we review the implementation of ESM/MUCA and BHM (including a
detailed description of the distinct dynamics adopted in this work). In
section \ref{nt:section} our numerical tests are presented and discussed.
Conclusions are in section \ref{conc:section}.

\section{The Multicanonical Method}
\label{es:section}

The idea of the Multicanonical method is to obtain the spectral
degeneracy of a given system using a biased RW in the configuration
space \cite{berg91,lee93}. The transition probability between two
states $X_{\rm old}$ and $X_{\rm new}$ is given by
\begin{equation}
\label{estransition}
\tau(X_{\rm old}, X_{\rm new}) = e^{-[S(E(X_{\rm new}))-S(E(X_{\rm
old}))]} = \frac{g(E_{\rm old})}{g(E_{\rm new})}
\end{equation}
where $S(E(X))=\ln g(E)$ is the entropy and $E(X)$ is the energy of
state $X$. The transitional probability (\ref{estransition})
satisfies a detailed balance equation and leads to a distribution of
probabilities where a state is sampled with probability $\propto
1/g(E)$. The successive visitations along the energy axis follow a
uniform distribution.  However, $g(E)$ is not known, {\em a priori}.
In order to obtain $g(E)$, Lee \cite{lee93} proposes the following
algorithm:

{\em Step 1:} Start with $S(E)=0$ for all states;

{\em Step 2:} Perform a few unbiased RW steps in the configuration space and
store $T(E)$, the number of tossed movements to each energy $E$
(in this stage, $T(E) = H(E)$ because all movements are accepted);

{\em Step 3:} Update $S(E)$ according to

\begin{equation}
\label{esentropy}
S(E) = \left\{ \begin{array}{ll}
                   S(E) + \ln T(E) &\mbox{, if $T(E) \neq 0$} \\
                   S(E)            &\mbox{, otherwise.}
               \end{array}
       \right. 
\end{equation}

{\em Step 4:} Perform a much longer MC run using the transitional probability
given by eq. (\ref{estransition}), storing $T(E)$.

{\em Step 5:} Repeat 3 and 4. This is considered one iteration.

This implementation is known to be quite sensitive to the lengths of
the MC runs in steps 2 and 4. In section \ref{nt:section} we study, in two
examples, how the accuracy depends on the total number and size of each
iteration.

\section{The Broad Histogram Method}
\label{bh:section}

BHM \cite{pmco96} enables us to directly calculate the energy
spectrum $g(E)$, without any need for a particular choice of the
dynamics to be used \cite{pmco98a}. Many distinct dynamic rules could
be used, and indeed some were already tested
\cite{pmco96,pmco98a,pmco98b,pmco98c,pmco98d,pmco99,wang98,munoz98,lima99,wang99,promb,j1j2}.

Within BHM, the energy degeneracy is calculated through the following steps
(alternatively, other quantities could replace $E$):

{\em Step 1}: Choice of a reversible protocol of allowed movements in the
state space. Reversible means simply that for each allowed movement $X_{\rm
  old}\rightarrow X_{\rm new}$ the back movement $X_{\rm new} \rightarrow
X_{\rm old}$ is also allowed. It is important to note that these movements are
virtual, since they are not actually performed.  In this work we take the
flips of one single spin as the protocol of movements;

{\em Step 2}: For a configuration $X$, to compute $N(X,\Delta E)$,
the number of possible movements that change the energy $E(X)$ by a
given amount $\Delta E$. Therefore $g(E) \langle N(E, \Delta
E)\rangle$ is the total number of movements between energy levels $E$
and $E+\Delta E$, according to the definition (\ref{microcanonical})
of microcanonical averages;

{\em Step 3}: Since the total number of possible movements from level
$E+\Delta E$ to level $E$ is equal to the total number of possible movements
from level $E$ to level $E+\Delta E$ (step 1, above), we can
write down the equation \cite{pmco96}
\begin{equation}
\label{bhrel}
g(E)\langle N(E, \Delta E)\rangle   =  g(E+\Delta E)\langle
N(E+\Delta E, -\Delta E)\rangle  .
\end{equation}
The relation above is exact for any statistical model and energy spectrum
\cite{pmco98c}. It can be rewritten as
\begin{equation}
\ln  g(E+\Delta E) -   \ln g(E) =  \ln \frac{\langle  N(E, \Delta
  E)\rangle}{\langle N(E+\Delta E, -\Delta E)\rangle}
\end{equation}
This equation can be easily solved for all values of $E$, after $\langle
N(E,\Delta E) \rangle$ is obtained by any procedure, determining $g(E)$ along
the whole energy axis. In cases where $\Delta E$ can assume more than one
value, eq.~(\ref{bhrel}) becomes an overdetermined system of equations.
However, the spectral degeneracy can be obtained without need of solving all
equations simultaneously, since the spectral degeneracy is the same for all
values of $\Delta E$.

The exact Broad Histogram relation (\ref{bhrel}) is independent of the
procedure by which $\langle N(E, \Delta E)\rangle$ is obtained
\cite{pmco98a,pmco98b,pmco98c,pmco98d,pmco99,wang98}.  Therefore, virtually
any procedure can be adopted in this task, for instance, an unbiased energy RW
\cite{pmco96}, a microcanonical simulation \cite{pmco98d}, or a mixture of
both \cite{pmco98c,pmco99}. Even the juxtaposition of histograms obtained
through canonical simulations at different temperatures, a completely
unphysical procedure, could be used, as in some of the results presented in
\cite{pmco98a}, and explicitly used in \cite{wang99} where BHM is
re-formulated under a transition matrix \cite{smith} approach. Here, we are
going to introduce an alternative procedure, referred as Entropic
Sampling-based Dynamics for BHM (ESDYN, hereafter). First, one implements ESM,
as described in section II, in order to perform the visitation in the
configuration space. Additionally, for each visited state $X$, we store the
values of $N(X, \Delta E)$ cumulatively into $E-$histograms. Therefore, at the
end we have two choices for the determination of the spectral degeneracy,
either by using the entropy accumulated through $T(E)$ (that is the
traditional ESM/MUCA) or by using the accumulated $\langle N(E, \Delta E)
\rangle$ and the BHM relation, eq.(\ref{bhrel}). Because of this special
implementation, we can guarantee that exactly the same states are visited for
both methods. Hence, the eventual difference in the performances reported in
this work must be credited to the methods themselves and not to purely
statistical factors.

The dynamic rule originally used in order to test BHM \cite{pmco96}
prescribes an acceptance probability $p = \langle N(E+\Delta
E,-\Delta E)\rangle / \langle N(E,\Delta E)\rangle$. Both the
numerator and the denominator are read from the currently accumulated
histograms, and thus $p$ varies during the simulation. Wang
\cite{wang98} has proposed a new approach: instead of using the
dynamically updated values of $\langle N(E, \Delta E) \rangle$ as in
\cite{pmco96}, the transitional probabilities follow a previously
obtained (from a canonical simulation, for example) distribution
$\langle N_{\rm fixed}(E,\Delta E )\rangle$, kept fixed during the
simulation. An alternative and simpler derivation of Wang's dynamics
can be done by using the BHM relation (\ref{bhrel}) itself.
>From this, we readily obtain that Wang's dynamics is the same as
using the transitional probability $p = g_{\rm fixed}(E)/g_{\rm
fixed}(E+\Delta E)$ with approximated values $g_{\rm fixed}(E)$
kept fixed during the simulation. We refer to this dynamics as
approximated Wang's dynamics, since $\langle N_{\rm fixed}(E,\Delta
E)\rangle$ (or $g_{\rm fixed}(E)$) is actually only an approximation
of the real $\langle N(E,\Delta E)\rangle$ (or $g(E)$). We will use
the results obtained by ESM or BHM with ESDYN as inputs to the
approximate Wang's dynamics. These dynamics will be called AWANG1 and
AWANG2, respectively.

For comparison, we also implemented a dynamics that uses the ESM
probabilities taken from the exact values of $g(E)$. It is worth
noticing that, as pointed out in the previous paragraph, this is
equivalent to Wang's proposal with exact values for $\langle N_{\rm
fixed}(E,\Delta E)\rangle$. We refer to this dynamics as WANG.  Its
purpose is only to test the accuracies of the other approaches, once
one does not know the exact $g(E)$ a priori, in real implementations.

\section{Numerical Tests}
\label{nt:section}

We start our comparison with the smallest system, since it is also
present in the ESM original paper by Lee \cite{lee93}. The partition
function for the $4 \times 4 \times 4$ simple cubic Ising model is
exactly known \cite{pearson82}.  It is given by the polynomial
function
\begin{equation}
Z(\beta)=\sum_{n=0}^{96}{C(n)u^n}
\end{equation}
where $u=\exp(-4 \beta )$, $\beta = 1/T$, $T$ is the temperature. The
energy spectrum is written in terms of the coefficients $C(n)$ as
$g(E)=g(2n)=C(n)$ for $n=0$ to $n=96$. For this model, only the first
$49$ coefficients are necessary by symmetry. The other $48$
coefficients are mirror images of the first ones.  Our results will
be expressed in terms of $S(E)=\ln g(E)$.

In order to compare the entropies as obtained by both ESM and
BHM, we normalize the entropy such that $g(96)=1$, i.e. $S(96)=0$.
This point corresponds to the center of the energy
spectrum, or, alternatively, to infinite temperature. Of course, the
error relative to the exact value vanishes for $E=96$. In fig. (1), we
compare the normalized (with respect to its exact value) entropy as
function of $E$ obtained by ESM and BHM (with four different
dynamics). In AWANG1 and AWANG2 dynamics we use as $\langle N_{\rm
fixed}(E,\Delta E )\rangle$ the results obtained by ESM and BHM with
ESDYN, respectively. BHM with the entropic
sampling dynamics or using the exact relation proposed by Wang present
errors within the same order of magnitude, while pure ESM gives the
worst results, as clearly seen in the inset. It is also clear that
AWANG1 and AWANG2 results are worse than BHM with ESDYN.

        The methods can be better compared by the ratio of their
relative errors rather than their absolute values. We define the
relative errors in the entropy, for a given energy $E$, as
\begin{equation}
\epsilon(E) = \left\vert
              \frac{ S(E)-S(E)_{\rm exact} }{S(E)_{\rm exact}}
              \right\vert
\end{equation}

In fig. (2) we show the ratio between the BHM relative errors (obtained by
ESDYN, AWANG1 and AWANG2 dynamics) and the ESM ones. The inset shows the
ratio between the errors from BHM with WANG and the ESM ones. In its worst
performance, the error obtained by BHM with ESDYN is roughly one third of
the one obtained by ESM. BHM with Wang's exact dynamics gives slightly
better results once it uses the exact $g(E)$ as input in order to get $g(E)$
as output. BHM with ESDYN gives on average results $11$ times more
accurate than ESM, for
this number of iterations. However, BHM with the AWANG1 and AWANG2
dynamics give relatively poorer results. Therefore, we have shown that the
ES dynamics is a powerful approach (among other possibilities
\cite{pmco96,pmco98a,pmco98b,pmco98c,pmco98d,pmco99,wang98,munoz98,lima99,wang99})
to obtain with great accuracy the microcanonical averages
$\langle N(E,\Delta E )\rangle$ needed by BHM. Let us stress that the
results for
$g(E)$ obtained with AWANG2 are worse than the input they use, namely the
values of $g(E)$ obtained as output of BHM with ESDYN. In order to obtain good
results with Wang's dynamics, we need a pretty good estimative of $g(E)$ as
input, and not only some crude estimation as claimed in \cite{wang98}.

        In fig. (3), we show the time evolution of the mean error as
a function of the number of Monte Carlo steps (MCS). One iteration in
ESM corresponds to perform many RW steps, according to
eq.~(\ref{estransition}), storing the number of visits at each energy
and, after this fixed number of RW steps, to update the entropy,
according to eq.~(\ref{esentropy}). As we pointed out before, the ESM
performance is quite sensitive to the choice of the number of RW
steps before each entropy update. For a fixed number of MCS ($10^6$), we
plot the time evolution for both ESM and BHM, but with different
number of iterations. As one can see in fig.(3), the best results
for ESM correspond to the smaller number of iterations since, in
this case, more RW steps are performed and, consequently, we have
better statistics for the determination of the entropy.

It is worth noticing that the ESM/MUCA errors seem to stabilize after a
few iterations. We believe that this effect occurs because the number of
visits to each energy is not sufficient to provide a good statistics in
determining the entropy. The only way to improve the accuracy in the
spectral degeneracy is to perform more RW steps between each entropy
update (it does not increase the CPU time if the total number of MCS is
kept constant). Conversely, for BHM the error decreases monotonically
because macroscopic quantities are stored, leading to a good statistics
even if some states are not frequently visited.  After the very first
steps, the errors decay as $t^{-1/2}$, as expected. It is also remarkable
the accuracy when using BHM: we reach the same accuracy of the best
performance of ESM/MUCA, by performing roughly 30 times less MCS.

Of course, accuracy is not the only important factor concerning the efficiency
of a computational method. BHM with ES dynamics has one additional step
compared to the traditional implementation of ESM, namely the storage of
the macroscopic quantities $N(X,\Delta E)$. So, we need to know the cost
of this additional step.  In table I, we show the CPU time (in seconds)
spent in both implementations on a 433 MHz DEC Alpha, in order to obtain
the results shown in fig.(1).  For direct comparison, we also present the
CPU time relative to the ESM CPU time. All dynamics tested within BHM
takes roughly the same CPU time. As one can see, BHM uses twice more CPU
time than ESM. However, for the same number of steps, and the best
strategy for ESM, BHM is at least 10 times more accurate.  If we consider
accuracy and CPU use, we argue that BHM is more efficient than ESM.

Up to now, we have tested both approaches in a very small lattice.
Nevertheless, we can show that BHM is even more efficient for larger
systems, as expected due to the macroscopic character of the
quantities $N(X,\Delta E)$. For $L^d$ Ising spins on a lattice, for
instance, even restricting the allowed movements only to single-spin
flips, the total number of movements starting from $X$ is just $L^d$.
Thus, being a finite fraction of them, $N(X,\Delta E)$ is a
macroscopic quantity (this is true along the whole energy axis,
except at the ground state where $g(E)$ presents a macroscopic jump
relative to the neighboring energy levels). In fig. (4) we present
results for the time evolution of the mean error for a $32 \times 32$
square lattice Ising model (a lattice that is 16 times larger than
the one in the previous results). The exact solution for this system
is also known \cite{beale96}. Here, the accuracy of BHM is two orders
of magnitude higher than that of ESM. Again the errors decrease as
$t^{-1/2}$ for BHM with ESDYN, while the ESM mean errors seem to
stabilize. In summary, BHM with ESDYN can obtain accurate results for
lattices much larger than the ones considered as limit for ESM.

\section{Conclusions}
\label{conc:section}

        The Multicanonical \cite{berg91,lee93} and the Broad
Histogram \cite{pmco96} methods are completely distinct from canonical
Monte Carlo methods, once they focus on the determination of the
energy spectrum degeneracy $g(E)$. This quantity is independent of
thermodynamic concepts and depends only on the particular system under
study. It does not depend on the interactions of the system with the
environment. Thus, once one has determined $g(E)$, the effects of
different environments can be studied using always the same data for
$g(E)$. Different temperatures, for instance, can be studied without
need of a new computer run for each $T$. 

The goal of this paper is to discuss the conceptual differences
between Multicanonical and Broad Histogram frameworks, and to compare
both methods concerning accuracy and speed. We obtained the energy
spectrum of the Ising model in $4 \times 4 \times 4$ and $32 \times
32$ lattices, for which the exact results are known and, therefore,
provide a good basis for comparison.  Our findings show that a
combination of the Broad Histogram method and the Entropic Sampling
random walk dynamics (BHM with ESDYN) gives very accurate results
and, in addition, it needs much less Monte Carlo steps to obtain the
same accuracy as the pure Entropic Sampling method. This advantage of
the Broad Histogram method grows with the system size, and it does
not present the limitations of the Multicanonical or Entropic
Sampling methods concerning large systems.

        The reason for the better performance is that the BHM
\cite{pmco96,pmco98a,pmco98b,pmco98c,pmco98d,pmco99,wang98,munoz98,lima99,wang99,promb,j1j2}
uses the microcanonical averages $\langle N(E,\Delta E) \rangle$
\cite{pmco96} of the macroscopic quantity $N(X,\Delta E)$ - the number of
potential movements which could be done starting from the current
state $X$, leading to an energy variation of $\Delta E$. In
this way, each new visited state contributes with a macroscopic value
for the averages one measures during the computer simulation. Being
macroscopic quantities, the larger the system, the more accurate are
the results for these averages. Conversely, Histogram
\cite{salzburg59,dickman84,swendsen93} and Multicanonical
\cite{berg91,lee93,hesselbo95} approaches rely exclusively on $H(E)$,
the number of visits to each energy. Therefore, each new averaging
state contributes with only one more count to the averages being
measured, i.e., $H(E)\rightarrow H(E)+1$, independent of the system
size.

Under a conceptual point of view, BHM is also completely distinct from
the other methods which are based on the final distribution of visits
$H(E)$. Alternatively, it is based on the determination of microcanonical,
fixed-$E$ averages $\langle N(E,\Delta E )\rangle$ \cite{pmco96},
concerning each energy level separately. Thus, the relative
frequency of visitation between distinct energy levels, which is
sensitive to the particular dynamic rule one adopts, i.e. the comparison
between $H(E)$ and $H(E')$, does not matter. The only requirement for the
dynamics is to provide a uniform sampling probability for the states
belonging to the same energy level. The transition probabilities from one
level to the others are irrelevant.

\section{Acknowledgments}

        This work has been partially supported by Brazilian agencies
CAPES, CNPq and FAPERJ. The authors acknowledge D. C.~Marcucci and J.
S. S\'a Martins for suggestions, discussions and critical readings of
the manuscript.

\begin{table*}[!h]
  \label{table:cputime}
  \begin{center}
    \begin{tabular}{|c|c|c|c|c|}
      \hline
      Method & Dynamics & {\begin{tabular}{c} CPU \\ Time(s)
        \end{tabular}}  & $\frac{t}{t_{\rm ESM}}$ &
      $\frac{1}{\langle\epsilon(E)/\epsilon(E_{\rm ESM})\rangle}$\\
      \hline
      ESM & ESDYN  & 4181 & 1    &  1     \\
      \hline
      BHM & AWANG1 & 9360 & 2.24 &  2.94  \\
      & AWANG2 & 9772 & 2.33 &  4.75  \\
      & ESDYN  & 8754 & 2.09 & 10.68  \\ 
      \hline
    \end{tabular}
  \end{center}
  \caption{CPU time for each method and dynamics. ESM/MUCA is at least
    twice faster than BHM with the various dynamics used in this work.
    However, BHM is more than 10 times more accurate than the
    multicanonical approach. For AWANG1 dynamics, one must add the
    previous ESM time spent in determining $\langle N_{\rm
    fixed}(E,\Delta E )\rangle$, i.e. 13541s. Analogously, AWANG2 spent a
    total time of 18526s. The simulations were carried out on a 433MHz DEC
    Alpha.  }
\end{table*}

\centerline{FIGURE CAPTIONS}

\baselineskip=22pt

\vspace{1cm}

\begin{itemize}

\item[FIG. 1 - ] Entropies (normalized by its exact values) for the
  $4 \times 4 \times 4$ Ising ferromagnet, obtained
  by ESM/MUCA and BHM for $100$ iterations of $10^6$ Monte Carlo steps,
  each.  The inset shows a detailed view of the first fourth of the whole
  spectrum. ESM/MUCA gives the largest errors. AWANG1 and AWANG2
  dynamics also give worse results than BHM with both ES or WANG dynamics.

\item[FIG. 2 - ] Ratio between the relative errors for BHM and
  the ESM/MUCA ones. The horizontal lines are the mean relative errors.
  BHM is on average more than ten times more accurate than
  ESM/MUCA, for this number of iterations. The exact Wang's dynamics
  is, as expected, slightly more accurate than ESDYN (as shown in
  the inset) once it uses the exact $g(E)$ as input in order to get
  $g(E)$ as output.  AWANG1 and AWANG2 give the worst results among
  the four distinct dynamics presently used in order to test BHM
  (see text).
  
\item[FIG. 3 - ] Time evolution of the mean error for BHM with ESDYN (a)
  and ESM/MUCA (b). In both cases, exactly the same averaging states are
  visited: thus, the differences are due to the methods themselves,
  not to statistics.  As described in the text, the entropic sampling
  dynamics is quite sensitive to the number of Monte Carlo steps
  between each iteration, that correspond to an update of the entropy.
  Here, we present the results for $N=10,100,1000$ iterations, that
  means $10^5,10^4$ and $10^3$ Monte Carlo steps, respectively,
  between each iteration.  The more Monte Carlo steps, the smaller is
  the error for the entropic sampling. Conversely BHM does not
  seem to depend on the computational strategy adopted. Moreover, the
  errors seem to stabilize after some steps, in the ESM case. On the
  contrary, for BHM, to get a better accuracy is a simple matter of
  increasing the computer time, once the errors decay as $t^{-1/2}$.
  The results are averaged over ten realizations.
  
\item[FIG. 4 - ] Time evolution of the mean error for BHM with ESDYN (a)
  and ESM/MUCA (b) for the $32 \times 32$ square lattice Ising Model. Now
  we consider $10^7$ MC steps.  Again we present the results for
  $N=100,1000$ iterations, that means $10^5$ and $10^4$ Monte Carlo
  steps, respectively, between each iteration.  The results are
  averaged over ten realizations. The ratio between the accuracies of
  BHM and ESM/MUCA is even higher for large systems.
\end{itemize}

\pagestyle{empty}
\newpage
\begin{figure}
\centerline{\epsfig{file=comp1.epsi,height=20cm}}
\end{figure}
\newpage
\begin{figure}
\centerline{\epsfig{file=comp2.epsi,height=20cm}}
\end{figure}
\newpage
\begin{figure}
\centerline{\epsfig{file=comp3.epsi,height=20cm}}
\end{figure}
\newpage
\begin{figure}
\centerline{\epsfig{file=comp4.epsi,height=20cm}}
\end{figure}

\end{document}